\documentclass[aps,twocolumn,floatfix,superscriptaddress,pra]{revtex4-1}
\usepackage{amssymb}
\usepackage{graphicx}
\usepackage{dcolumn}
\usepackage{amsmath}
\usepackage{bm}
\usepackage{yhmath}
\usepackage{textcomp}
\usepackage{epstopdf}
\usepackage{color}
\usepackage{ulem}
\usepackage[toc,page,title,titletoc,header]{appendix}
\usepackage[colorlinks,
            linkcolor=blue,
            anchorcolor=blue,
            citecolor=blue
            ]{hyperref}
\usepackage{natbib}

\usepackage{tikz}
\newcommand*\emptycirc[1][0.4ex]{\tikz\draw (0,0) circle (#1);} 
\newcommand*\fullcirc[1][0.4ex]{\tikz\fill (0,0) circle (#1);}

\begin{document}
 \title{Lattice control of non-ergodicity in a polar lattice gas}
\author{H. Korbmacher}
\affiliation{Institut f\"ur Theoretische Physik, Leibniz Universit\"at Hannover, Germany}
\author{P. Sierant} 
\affiliation{ICFO-Institut de Ci\`encies Fot\`oniques, The Barcelona Institute of Science and Technology, Av. Carl Friedrich
Gauss 3, 08860 Castelldefels (Barcelona), Spain}
\author{W. Li}
\affiliation{Institut f\"ur Theoretische Physik, Leibniz Universit\"at Hannover, Germany}
\author{X. Deng}
\affiliation{Leibniz-Rechenzentrum, D-85748 Garching b. M\"unchen, Germany}
\author{J. Zakrzewski}
\affiliation{Instytut Fizyki Teoretycznej, 
Uniwersytet Jagiello\'nski,  \L{}ojasiewicza 11, PL-30-348 Krak\'ow, Poland}
\affiliation{Mark Kac Complex Systems Research Center, Uniwersytet Jagiello{\'n}ski, PL-30-348 Krak{\'o}w, Poland}
\author{L. Santos}
\affiliation{Institut f\"ur Theoretische Physik, Leibniz Universit\"at Hannover, Germany}

\begin{abstract}

Strong-enough inter-site interactions may result in lack of ergodicity in disorder-free many-body lattice systems.
Ultra cold dipolar gases in optical lattices provide an experimentally accessible platform for exploring this physics.
Dipolar inter-site interactions are usually assumed to decay with a fixed power-law. We show that 
in a one-dimensional polar lattice gas the actual decay depends on the transversal confinement.
This affects profoundly the particle dynamics, which mimics rather that of a system with an externally-controllable effective power-law interaction. Our results show that the crucial role of the interaction decay on disorder-free localization may be flexibly studied in experiments with polar gases.
\end{abstract} 
  
\maketitle



\section{Introduction}


Many-body localization~\cite{Nandkishore15, Alet18, Abanin19} constitutes a major exception to the thermalization paradigm in many-body systems. 
Although initially considered in the presence of disorder, recent years have witnessed a rapidly growing interest on non-ergodic disorder-free systems~\cite{Carleo12,Grover14,Schiulaz15,VanHorssen15, Barbiero15,Papic15,Hickey16,Smith17b,Mondaini17,Schulz19, vanNieuwenburg19, Taylor20, Chanda20c, Yao20b, Scherg21, Guo21, Morong21, Yao21a,Yao21}. 
Disorder-free localization occurs naturally due to dynamical constraints~\cite{Lan18, Feldmeier19, Nandkishore19}, which 
result in a finite number of conservation laws, inducing Hilbert space fragmentation~\cite{DeTomasi19,Pietracarpina19,Sala20,Khemani20, Herviou20, Yang20}. 

Ultra cold particles in optical lattices and tweezer arrays constitute an exceptional system for studying out-of-equilibrium many-body quantum systems~\cite{Langen2015}, as recently highlighted by experiments on many-body localization~\cite{Schreiber15, Choi16, Rispoli19, Lukin19}, quantum scars~\cite{Bernien17}, and 
(disorder-free) Stark localization~\cite{Scherg21, Morong21}. 
Most current lattice experiments involve contact-interacting particles.   
In the tight-binding regime, those experiments simulate different forms of the Hubbard model with on-site interactions~\cite{Jaksch98, Greiner02}, although 
weak nearest-neighbor~(NN) interactions may result from super-exchange~\cite{Trotzky08}.

Recent experiments are focusing on long-range interacting lattice systems, 
including trapped ions~\cite{Richerme14, Jurcevic14}, Rydberg gases~\cite{Bernien17,Browaeys20,Guardado-Sanchez21,Scholl2021}, 
and polar lattice gases of magnetic atoms~\cite{DePaz2013, Baier2016, Patscheider2020}, and polar molecules~\cite{Yan2013}.
These gases are characterized by strong inter-site interactions, and hence allow for the realization of different spin models and extended Hubbard models~(EHMs) \cite{Dutta15}. 
Spin models have been realized in magnetic atoms~\cite{DePaz2013, Patscheider2020}, polar molecules~\cite{Yan2013}, and Rydberg atoms~\cite{Scholl2021}, whereas 
seminal EHM experiments have been performed using magnetic~\cite{Baier2016} and Rydberg atoms~\cite{Guardado-Sanchez21}.
Inter-site interactions result in an intriguing dynamics in EHMs~\cite{Valiente09,Nguenang09,Petrosyan07,Li20,Morera21, Fukuhara13, Salerno20, Li21b}. 
In particular, the combination of energy conservation, finite band-width, and dipolar interactions is expected to result in Hilbert-space shattering and disorder-free localization for strong-enough dipolar interactions~\cite{Li2021}.

Long-range systems present inter-site interactions which may potentially extend well beyond NNs. In trapped ions, the power law interaction 
$1/r^\beta$~(with $r$ the intersite distance), may be externally tailored~($0<\beta<3$) using laser dressing~\cite{Richerme14, Jurcevic14}. In polar gases, due to the form of the dipolar interactions, the dipolar tail is typically assumed to decay with a fixed power law $1/r^3$. However, this assumption must be carefully reconsidered, especially in low-dimensional models, since inter-site dipolar interactions are affected by the geometry of the on-site Wannier functions~\cite{Sowinski12,Wall2013}. 

In this paper, we show that the dipolar tail acquires in one-dimensional~(1D) lattices a universal analytic dependence on the transversal confinement, which may depart under typical conditions strongly from the quite generally assumed in up-to-date studies, $1/r^3$ form. This leads to a very significant modification of the dynamics of 1D hard-core polar lattice gases, which, remarkably, mimics that of a model with variable power-law interactions, 
$1/r^{\beta\neq 3}$, where the power $\beta$, and with it the localization threshold, may be controlled by the transversal confinement. 
Our results show that near future experiments on polar gases may hence provide a surprisingly flexible 
platform for the study of the key role of inter-site interactions on disorder-free localization in many-body lattice systems.


The structure of the paper is as follows. In Sec.~\ref{sec:EHM} we discuss the model under consideration. Section~\ref{sec:Tail} is devoted to 
the dependence of the dipolar tail on the transversal confinement. 
In Sec.~\ref{sec:HSF} we analyze how 
Hilbert space fragmentation in the polar lattice gas is affected by the transversal confinement. Section~\ref{sec:DW} analyzes the particle 
dynamics when starting with an initial density wave, whereas Sec.~\ref{sec:General} discusses the case of general initial Fock state. 
In Sec.~\ref{sec:Experiments} we comment on possible experimental realizations,  while 
 Sec.~\ref{sec:Conclusions} summarizes our conclusions. More technical details are discussed in the appendices.



\section{Extended Hubbard model}
\label{sec:EHM}
We consider dipolar bosons of mass $m$ in a 1D optical lattice, $V_0\sin^2(\pi z/\lambda)$, transversally confined by an isotropic harmonic 
potential $\frac{1}{2}m\omega_\perp^2 (x^2+y^2)$. The dipole moments are assumed to be 
oriented by an external field on the $xz$ plane forming an angle $\alpha$ with the lattice axis $z$. 
For a sufficiently deep lattice, the system is well described by the EHM:
\begin{equation}
\hat H = -t\sum_{i} \left ( \hat b_i^\dag \hat b_{i+1}+\mathrm{H.c.} \right ) + \sum_i \sum_{j>0} V_j \hat n_{i} \hat n_{i+j},
\end{equation}
where $\hat b_i$~($\hat b_i^\dag$) is the annihilation~(creation) operator at site $i$, $\hat n_i = \hat b_i^\dag \hat b_i$, and we impose the hard-core 
constraint $( \hat b_i^\dag )^2=0$. This constraint is well justified if no site is multiply occupied initially, and if the on-site interactions are large-enough to prevent multiple occupations at any later time. Although the on-site interactions depend on the dipole-dipole interaction, for strong-enough short-range interactions (which may require the use of Feshbach resonances), we can neglect multiple occupation at any time for all values of the dipole strength considered. The hard-core constraint implies negligible collisionally-assisted hops, which may be relevant in the soft-core regime~\cite{Maik2013,Kraus2020}.

The inter-site interaction for dipoles $j$ sites apart is characterized by the coupling constant~\cite{exchange}:
\begin{equation}
V_j = \int d^3 r \int d^3r' V(\vec r-\vec r\,') |\varphi(\vec r)|^2 |\varphi(\vec r\,'-j\lambda \vec e_z)|^2
\label{eq:Integral}
\end{equation}
where $V(\vec r)=\frac{C_{dd}}{4\pi r^3} \left ( 1-3 \frac{\left ( x\sin\alpha + z\cos\alpha \right )^2}{r^2} \right)$ is the dipole-dipole interaction. For magnetic dipoles, 
$C_{dd}=\mu_0 \mu^2$, with $\mu_0$ the vacuum permeability and $\mu$ the magnetic moment. For electric dipoles, 
$C_{dd}=\frac{d^2}{\epsilon_0}$, with $\epsilon_0$ the vacuum dielectric constant, and $d$ the electric dipole moment.
We characterize below the dipole strength by the dipolar length, $a_{dd}=\frac{mC_{dd}}{12\pi\hbar^2}$. 
The on-site wave function, $\varphi(\vec r)=\phi_0(x,y)W(z)$, is given by the Wannier function $W(z)$  
associated to the lowest-energy band, and  by the ground-state of the transversal trap, 
$\phi_0(x,y)=\frac{e^{-(x^2+y^2)/2l_\perp^2}}{\sqrt{\pi} l_\perp}$, with 
$l_\perp^2=\hbar/m\omega_\perp$~(we assume that $\hbar\omega_\perp$ is much larger than other energies involved in the EHM).



\begin{figure} [t]
\begin{center}
\includegraphics[width=\columnwidth]{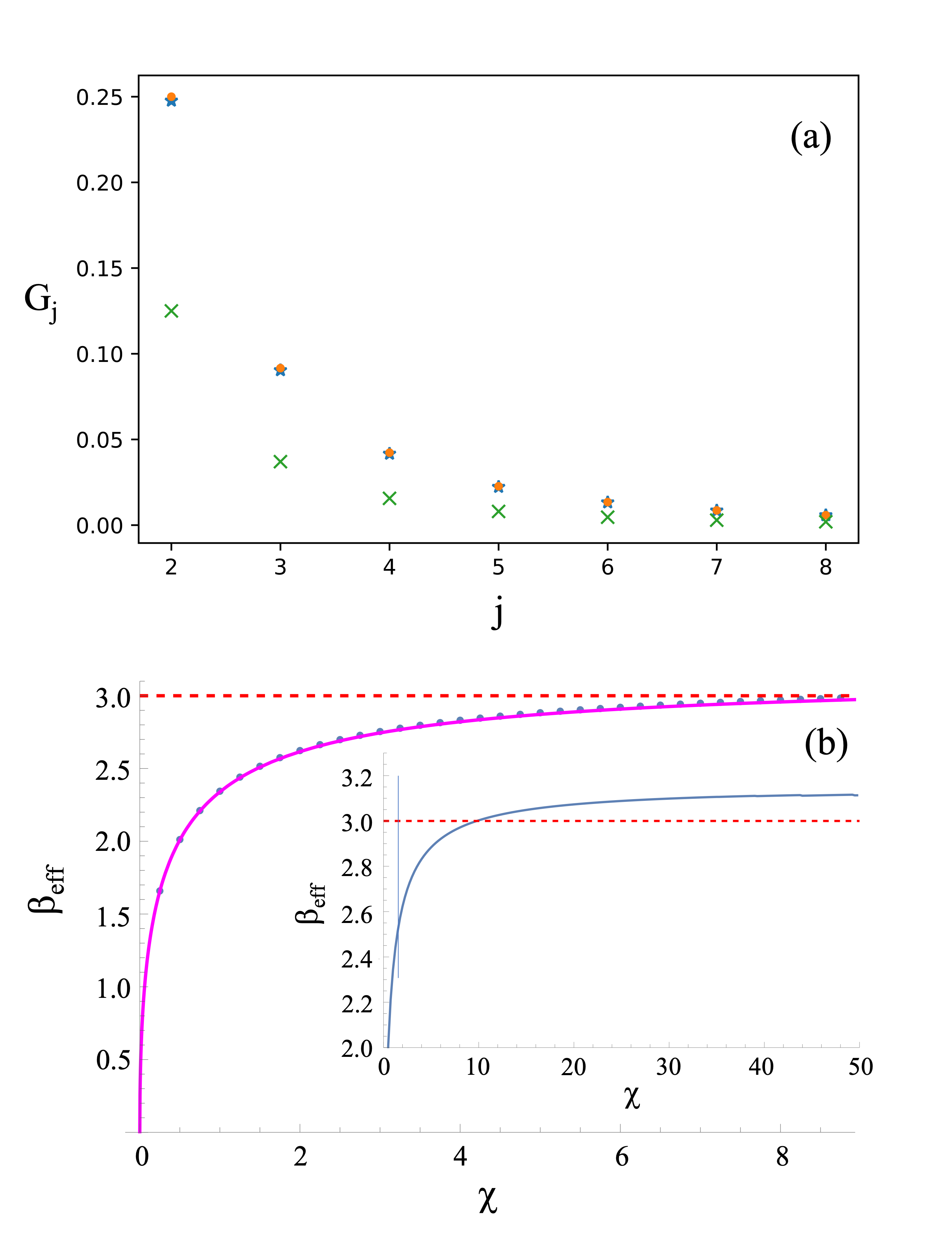}
\caption{(a) Dipolar tail $G_j$ for $s=8$ and $\chi=0.47$~($B=2.54$). Blue stars and orange circles depict, respectively, the results obtained directly using the exact Wannier functions, and using Eq.~\ref{eq:function-f}. Green crosses indicate the $1/j^3$ tail.
(b) Power $\beta_{\mathrm{eff}}$ for $s=30$ 
and different $\chi$. The dots indicate the results 
obtained employing Eq.~\eqref{eq:Integral} with the exact Wannier functions. The solid line depicts $-\log(G_2)/\log(2)$. 
The inset shows the results for a wider range of $\chi$ values. 
In both graphs the dashed line indicates $\beta_{\mathrm{eff}}=3$.}
 \label{fig:1}
\end{center}
\end{figure}


\section{The dipolar tail}  
\label{sec:Tail}
For deep-enough lattices, we may approximate $W(z)\simeq \frac{e^{-z^2/2l^2}}{\sqrt{\sqrt{\pi}l}}$, with $l=\frac{\lambda}{\pi s^{1/4}}$, where $s=\frac{V_0}{E_{\mathrm{R}}}$ and 
$E_{\mathrm{R}}=\frac{\pi^2\hbar^2}{2m\lambda^2}$  is the recoil energy.
We then obtain for $l_\perp>l$~(see App.~\ref{sec:App1}):
\begin{equation}
\frac{V_j}{E_{R}} =\frac{3B^{3/2}}{2\pi^2}\left ( 3\cos^2\alpha -1\right ) \left ( \frac{a_{dd}}{\lambda} \right )  f(\sqrt{B}j),
\label{eq:Vj}
\end{equation}
where $B=\frac{\pi^2}{2}\frac{\chi}{1-\frac{\chi}{2\sqrt{s}}}$, $\chi=\frac{\hbar\omega_\perp}{E_{\mathrm{R}}}$, and 
\begin{equation}
f(\xi)=2\xi-\sqrt{2\pi} (1+\xi^2)e^{\xi^2/2}\mathrm{erfc}\left ( \xi/\sqrt{2} \right ).
\label{eq:function-f}
\end{equation}
Denoting $V\equiv V_1$, we can write $V_j = V G_j(B)$, with $G_j(B)=f(\sqrt{B}j)/f(\sqrt{B})$~\cite{HA}. Hence, actual hard-core dipoles in 1D lattices have a universal dependence on both $V/t$ and $B$, that characterize, respectively, the dipole strength and the dipolar tail.

This tail must be compared to the $1/j^3$ decay, which is quite generally assumed in studies of polar lattice gases~(for a 
comparison for $B=2.54$ see Fig.~\ref{fig:1}(a)). Although the 
$1/j^3$ decay is eventually recovered at sufficiently long distances, i.e. $G_{j\to\infty}(B)\to 1/j^3$, the correction may be very relevant for the first nearest neighbors. The modification of the ratio $G_2$ between next-to-NN and NN interactions is particularly relevant, since this ratio is crucial for the Hilbert-space fragmentation and dynamics in a polar lattice gas. In contrast, as shown below, beyond next-to-NN interactions play a relatively minor role. Hence, we introduce at this point the effective power $\beta_{\mathrm{eff}}(B)$, see Fig.~\ref{fig:1}(b), such that $G_2(B)=1/2^{\beta_{\mathrm{eff}}(B)}$~\cite{Gj}. Note that there is a one-to-one correspondence between $B$ and $\beta_{\mathrm{eff}}(B)$, and hence the hard-core 
lattice gas will present universal properties in $V/t$ and $\beta_{\mathrm{eff}}$.

For $l>l_\perp$~($\chi>2\sqrt{s}$), 
$\beta_{\mathrm{eff}}(B)>3$, i.e. the dipolar tail decays 
to the first nearest neighbors faster than $1/j^3$. 
In that regime, for large-enough $\chi$ and $s$, $\beta_{\rm{eff}}$ 
approaches $\beta_\infty \simeq 3 + \frac{1}{\mathrm{ln} (2)} \frac{9}{2\pi^2  \sqrt{s}}$~(see App.~\ref{sec:App2}). As a result, 
for $l>l_\perp$, next-to-NN are less relevant than in the $1/j^3$ model, and the correction to the $1/j^3$ dependence due to the transversal confinement induces only minor modifications in the dynamics~(also the ground-state properties are only slightly affected~\cite{Wall2013}). 

In stark contrast, for $l<l_\perp$~($\chi<2\sqrt{s}$), $\beta_{\mathrm{eff}}(B)$ may become significantly smaller than $3$~(see Fig.~\ref{fig:1}~(b)), i.e. the dipolar tail decays significantly slower than $1/j^3$ for the first nearest neighbors. The markedly enhanced role of the next-to-NN interactions leads to a strongly modified dynamics, as shown below.


\section{Hilbert-space fragmentation}  
\label{sec:HSF}
For the model with just NN interactions~(NN model), $V_j=V\delta_{j,1}$, increasing $V/t$ results in 
an emerging dynamical constraint, given by the conservation of the number of NN bonds $N_{\mathrm{NN}}=\sum_j \langle n_j n_{j+1}\rangle$. This constraint
leads to Hilbert-space fragmentation into dynamically unconnected blocks~\cite{DeTomasi19}. 
The presence of a $1/j^3$ tail results for large-enough $V/t$
 in a strong fragmentation~(shattering) of the NN blocks due to the emerging 
 conservation of the number of next-to-NN bonds, $N_{\mathrm{NNN}}=\sum_j \langle n_j n_{j+2}\rangle$~\cite{Li2021}. 
 As shown below, the controllable modification of $V_2/V$ significantly affects this shattering, and 
 with it the particle dynamics.
 


\begin{figure} [t!]
\begin{center}
\includegraphics[width=\columnwidth]{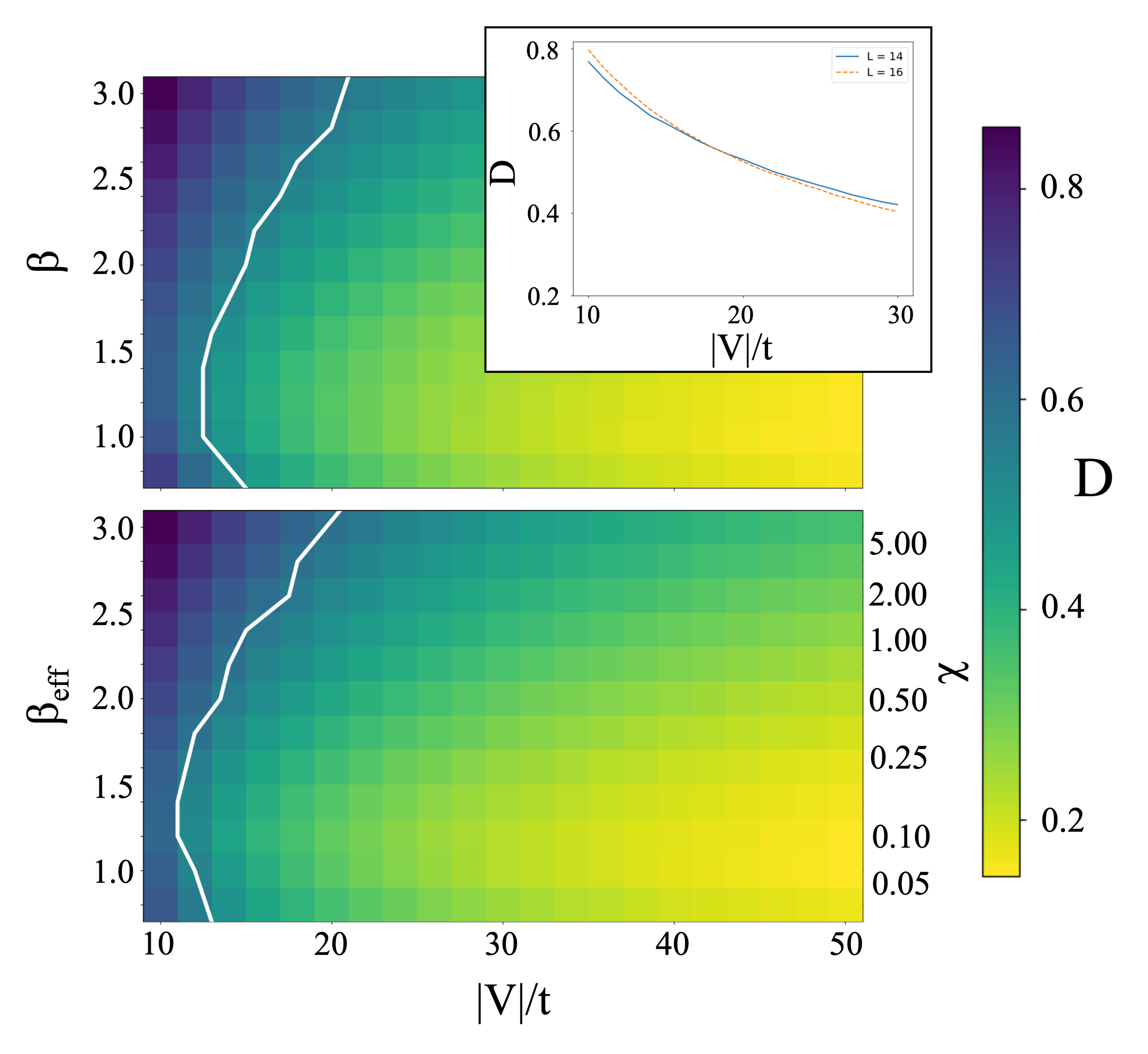}
\caption{Shattering of the NN blocks. (top) Average $D$ for $V/j^\beta$ interactions, as a function of $V/t$ and 
$\beta$. (bottom) Same for a polar lattice gas, as a function of $V/t$ and $\beta_{eff}$ (right axis) and $\chi$~(for $s=30$)~(left axis). The color plot is obtained for $N=8$ particles in $L=16$ sites with open boundary conditions. The inset shows $D$ as a function of $V/t$ for a dipolar gas with $B=11.5$~($\beta_{\mathrm{eff}}\simeq 2.6$) for $N=7$ and $L=14$ and  $N=8$ and $L=16$. The crossing point provides an estimation of the shattering transition. The white curves in both panels show the crossing points of the $D$ curves for $N=7$ and $L=14$ and $N=8$ and $L=16$. }
 \label{fig:2}
\end{center}
\end{figure}

 
In order to study Hilbert-space fragmentation, we employ 
exact diagonalization to obtain the eigenstates
$|\alpha\rangle$ of $N$ particles in $L$ sites with open boundary conditions.  
We then express the Fock states $|f\rangle = \prod_{l=1}^L  |n_l (f)\rangle$ with population $n_l(f)=0,1$ in site $l$, 
in the basis of eigenstates, $|f\rangle = \sum_\alpha \psi_f(\alpha)|\alpha\rangle$. For the NN model, a sufficiently large $V/t>10$ results in Hilbert space 
fragmentation into unconnected blocks~(NN blocks) with a size much smaller than the overall Hilbert space dimension. 
Further fragmentation of the NN blocks due to beyond-NN interactions 
is characterized for each Fock state $|f\rangle$ by the fractal dimension, $D_f=-\ln({\cal I}_f)/\ln(\Lambda_f)$~\cite{Mace2019}, 
where $\Lambda_f$ is the size of the NN block to which $|f\rangle$ belongs, and 
${\cal I}_f = \sum_{\alpha} |\psi_\alpha(f)|^{4}$ is the inverse participation ratio. $D_f$ approaches zero when the NN block shatters.
The average, $D$, of $D_f$ over the whole Fock basis provides a good quantitative estimation of the shattering.

Figure~\ref{fig:2} shows $D$, as a function of $V/t$, for (a) $V_j = 1/j^\beta$, and (b) $V_j=VG_j(B)$, characterized by $\beta_{\mathrm{eff}}(B)$. The comparison of both graphs shows that, due to the dominant role played by the next-to-NN interactions, Hilbert-space shattering in an actual dipolar gas may closely mimic that of a system with modified power-law interactions $1/j^{\beta=\beta_{\mathrm{eff}}(B)}$. Note also the potentially very large deviation from the results expected for a $1/j^3$ tail.



\begin{figure} [t!]
\begin{center}
\includegraphics[width=1.05\columnwidth]{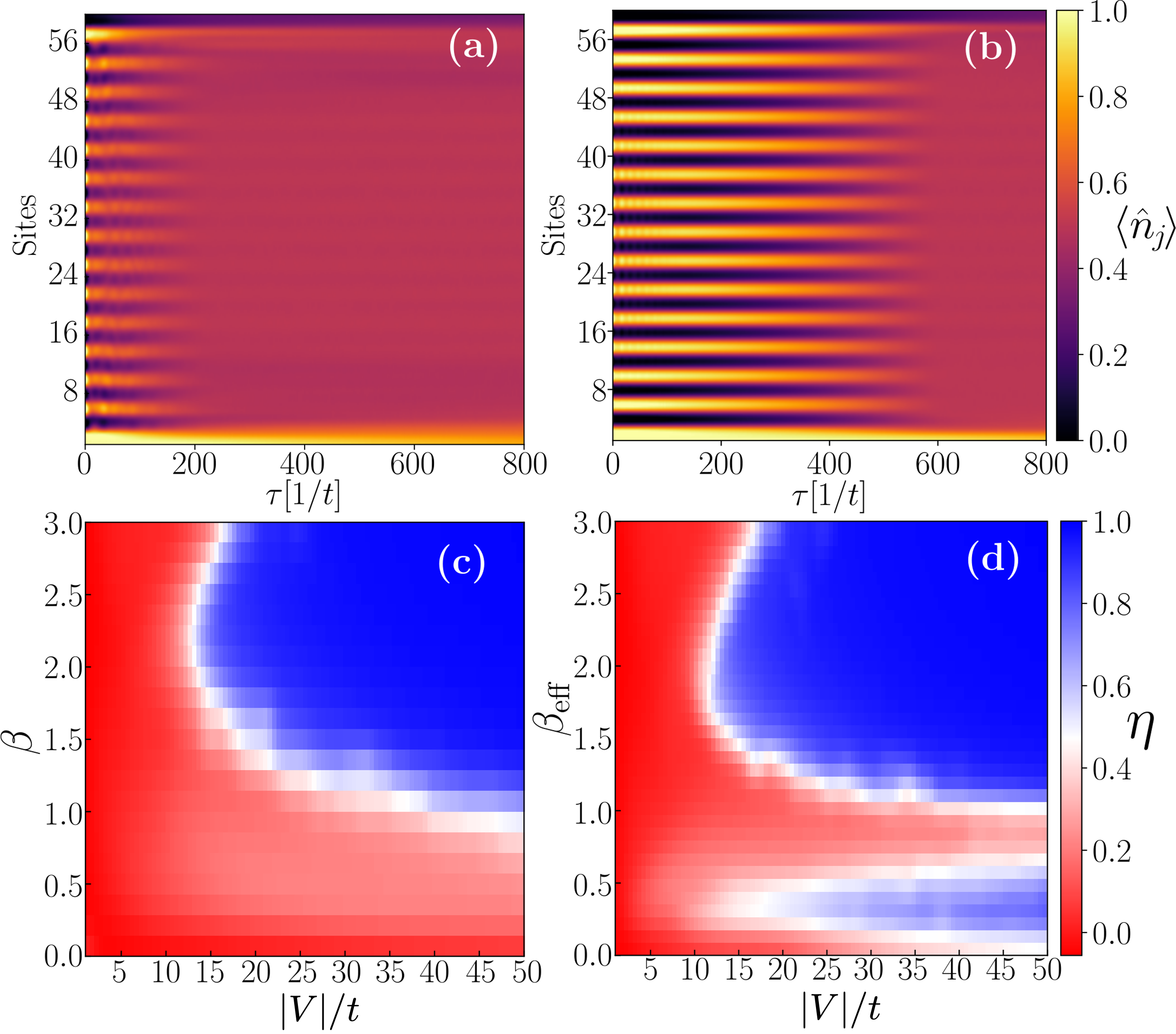}
\caption{Homogenization of an initial density wave. Average density at different sites as a function of  time for 
(a) $V_j=V/j^3$ and (b) $V_j=G_j(B)$, for 
$B=2.54$~($\beta_{\mathrm{eff}}\simeq 2$). In both cases $V/t=16$. Results obtained from Time-Dependent Variational Principle (TDVP) calculations for $L=60$ and $N=30$~(see App.~\ref{sec:TDVP}).  
Panels~(c) and~(d) show the inhomogeneity parameter averaged in the time interval $190/t<\tau<210/t$, evaluated for a system with $L=24$, $N=12$ and an initial density wave as in (a) and (b). Panel~(c) shows the results for a model $V/j^\beta$ as a function of $\beta$ and $V/t$, which should be compared with (d) where the results for a polar gas and different values of $\beta_{\mathrm{eff}}$ and $V/t$ are presented. The results were obtained using Chebyshev time propagation~(see App.~\ref{App:4}).}
 \label{fig:3}
\end{center}
\end{figure}



\section{Dynamics of an initial density wave}
\label{sec:DW}
The actual form of the interaction decay has relevant consequences for the dynamics of particles, well illustrated by the relatively simple case of an initial density wave:
$$
|\fullcirc\,\fullcirc\, \emptycirc\,\emptycirc\,\fullcirc\,\fullcirc\, \emptycirc\,\emptycirc\, \fullcirc\,\fullcirc\, \emptycirc\,\emptycirc\,\cdots\rangle,
$$
with open boundary conditions,
which may be prepared using a superlattice~(similar initial conditions have been recently 
studied in Rydberg gases~\cite{Guardado-Sanchez21}).  
Figure~\ref{fig:3} compares the dynamics for a 
$1/j^3$ decay, and the actual evolution for $B=2.54$~($\beta_{\mathrm{eff}}\simeq 2$) for $V/t=16$. 
Although in both cases we observe delocalization at long times, the homogenization is approximately four times slower in the actual polar lattice gas. 
Density homogenization is well characterized by the inhomogeneity parameter, 
\begin{equation}
\eta = \frac{1}{2L\rho(1-\rho)} \sum_{j=1}^L \left |\langle \hat n_j \rangle - \rho \right |,
\end{equation}
with $\rho=N/L$. Note that $\eta$ ranges from 1 for a maximally inhomogeneous state (i.e. for a Fock state), to $0$ for a fully homogeneous density, with $\langle \hat n_i \rangle =\rho$, for all sites $i$. Figures~\ref{fig:3}~(c,d) show $\eta$, after a time $\tau=200/t$ for, respectively, $V_j=V/j^\beta$ and 
a polar gas with different $\chi$~(and hence different $\beta_{\mathrm{eff}}$). 
In both cases a marked jump in $\eta$ as a function of $V/t$ characterizes the 
onset of strong localization. Note as well the remarkable similarity down to $\beta_{\mathrm{eff}}\simeq 1$~($B\simeq 0.3$), of both 
graphs as a function of, respectively, $\beta$ and $\beta_{\mathrm{eff}}$. Hence, for $B>0.3$, the dynamics and the localization threshold at a given time in a polar lattice gas are basically indistinguishable from those in a system with power-law interactions $1/j^{\beta_{\mathrm{eff}}}$. 

For $\beta_{\mathrm{eff}}<1$, the dynamics of a polar lattice gas departs significantly 
from that of a $1/j^{\beta_{\mathrm{eff}}}$ model~(see Fig.~\ref{fig:3}). In particular, the polar gas presents a marked resonance for $\beta_{\mathrm{eff}}\simeq 0.8$~($B\simeq 0.17$), at which the 
gas becomes quickly homogeneous even for large $V/t$. Such a resonance is absent in the corresponding power-law model, which presents a steady re-entrance of 
the delocalized regime. This is easy to understand, since for $\beta\to 0$, the inter-site interactions become a constant of motion, 
$\frac{V}{2}\sum_{i\neq j} n_i n_j = V N(N-1)/2$, with $N$ the total particle number, and hence the system is formed effectively by non-interacting hard-core bosons. 
This steady growth of the extended regime is not present in the actual polar lattice gas, because for long distances $G_j \to 1/j^3$. Only for $\beta_{\mathrm {eff}}\to 0$ delocalization 
extends to large $V/t$ values.


\section{Dynamics for general initial Fock states.--}
\label{sec:General}
\begin{figure} [t!]
\begin{center}
\vspace{-0.2cm}
\includegraphics[width=\columnwidth]{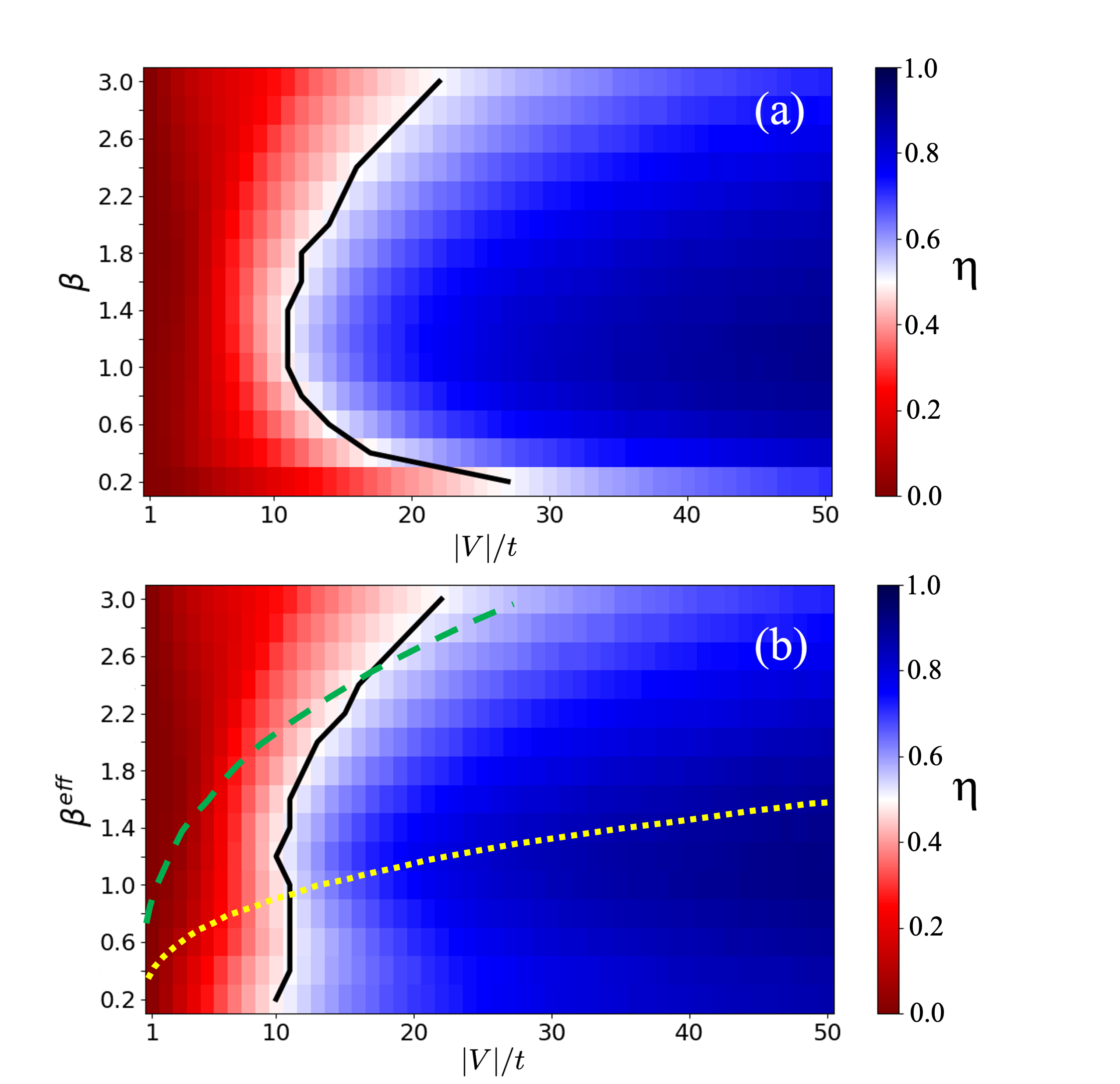}
\caption{Inhomogeneity $\eta$ after a time $\tau=200/t$, evaluated for a system with $L=16$ sites, and averaged over all initial Fock states with 
$N=8$ and initial $N_{NN}=4$. Panels~(a) and (b) show, respectively, the case of $V_j=V//j^\beta$ and of a polar lattice gas. 
The thick black curve indicates the line with $\eta=0.5$. 
The results were obtained using exact time evolution~\cite{tenpy}.
The dashed green~(dotted yellow) curves indicate the 
values in the ($V/t$,$\beta_{\mathrm{eff}})$ space achieved for different $\omega_\perp$ for $^{164}$Dy in an UV lattice of $\lambda=180\,\mathrm{nm}$~(for NaK, with $d=0.8\,$Debye, in a lattice with $\lambda=500\,\mathrm{nm}$) with a fixed lattice depth $s=8$. 
For the dashed-green~(dotted-yellow) curve, from left to right, $\chi$ varies from $0.0167$ to $4.68$~(from $0.0045$ to $0.21$).}
\label{fig:5}
\end{center}
\end{figure}

The previous conclusions are, for $\beta_{\mathrm{eff}}>1$, largely
representative of the dynamics for more general initial Fock 
states. We have evaluated the dynamics of all possible initial Fock states for $L=16$ and open boundary conditions, fixing only $N=8$ and an initial $N_{NN}=4$. Note that this set includes $4410$ states with different number of clusters of various lengths. Figure~\ref{fig:5} shows for $V_j=V/j^\beta$ and for a polar lattice gas, the value of $\eta$ (averaged over all possible initial conditions) after an evolution time $\tau=200/t$. 
Also for this more general case, there is a marked transition between localization and delocalization. However, the behavior of the averaged $\eta$ is less abrupt, due to the 
difference in the degree of localization between different initial conditions. The results for both models are again remarkably similar, down to $\beta_{\mathrm{eff}}\simeq 1$~($B>0.3$). For lower $\beta_{\mathrm{eff}}$ the results are markedly different. Note the absence of a resonance in the polar gas, which 
is a specific feature of an initial density wave. Note that also for general initial conditions, the power-law model presents at low $\beta$ a marked re-entrance of the 
extended regime, absent in polar lattice gases at low $B$.



\section{Experimental relevance} 
\label{sec:Experiments}
Our results are directly relevant for on-going and near-future experiments with low-dimensional dipolar gases in optical lattices, including magnetic atoms, polar molecules, and Rydberg gases. For a polar lattice gas, our analysis reveals that there should be a potentially very significant difference between the case in which one-dimensionality is achieved in a 3D lattice, and the case in which it is obtained by means of transversal harmonic confinement.  In the former case, the suppression of transversal hopping requires a very strong lattice on the transversal directions, resulting in $l_\perp \ll l$. The typically assumed $1/j^3$ dependence is hence approximately reached for a sufficiently large lattice depth $s$~\cite{Wall2013}. In the latter case, in contrast, even tight transversal confinements may result in strongly modified properties. 
For example, a lattice depth of $s=20$ and a transversal confinement with $\chi\simeq 0.5$ results in $\beta_{\mathrm{eff}}(B)\simeq 2$. For $^{164}$Dy in an UV lattice with 
$\lambda=180$nm this would require $\omega_\perp/2\pi \simeq 4.6\,$kHz, whereas a frequency $1.58\,$kHz will be needed for a NaK molecule in a lattice with $\lambda=500$nm. 
Note that changing $\omega_\perp$, 
modifies both $V/t$ and $\beta_{\mathrm{eff}}$. 
In Fig.~\ref{fig:5}(b) we show the trajectories in $(V/t,\beta_{\mathrm{eff}})$ for $s=8$ obtained for different $\chi$ for Dy and NaK (for $s=8$ the hopping times are respectively $10\,\mathrm{ms}$ and $31\,\mathrm{ms}$). Note that in both cases the localization-to-delocalization transition may be crossed by changing $\omega_\perp$ at a fixed lattice depth.



\section{Conclusions}
\label{sec:Conclusions}
One-dimensional polar lattice gases are characterized by inter-site interactions that decay following a universal dependence on  
the transversal confinement and the lattice depth. This decay, which may depart very significantly from the quite generally assumed $1/r^3$ form, 
strongly affects the dynamics of hard-core systems. Interestingly, due to the dominant role played by nearest-neighbor and next-to-nearest-neighbor interactions, both Hilbert-space fragmentation and homogenization dynamics become basically identical to a model with an externally-controllable power-law decay. 
As a result, polar lattice gases constitute a flexible platform for the study of the role of 
inter-site interaction in disorder-free 
many-body localization. Similarly, one may anticipate that, in the presence of disorder, the  critical disorder amplitude leading to the 
extended to localized crossover may depend non-trivially on the interactions tail. Such a study, as well as the treatment of soft-core bosons \cite{adith22}, will be discussed elsewhere.



\begin{acknowledgments}
J.Z. thanks Titas Chanda for help with TDVP implementation.
We acknowledge support of the Deutsche Forschungsgemeinschaft (DFG, German Research Foundation) under Germany's Excellence Strategy -- EXC-2123 QuantumFrontiers -- 390837967, 
and FOR 2247.  Some of the numerical computations have been possible thanks to PL-Grid Infrastructure. The work of  J.Z. has been realized within the Opus grant 2021/43/I/ST3/0114, financed by National Science Centre (Poland). X.D. aknowledges support of BMBF through DAQC.
P.S. acknowledges support from Ministerio de Ciencia e Innovaci\'on, Agencia Estatal de Investigaciones (R\&D project CEX2019-000910-S, AEI/10.13039/501100011033, Plan National FIDEUA PID2019-106901GB-I00, FPI), Fundació Privada Cellex, Fundació Mir-Puig, and from Generalitat de Catalunya (AGAUR Grant No. 2017 SGR 1341, CERCA program)
\end{acknowledgments}



\begin{thebibliography}{81}
\bibitem [{HA()}]{HA}%
  \BibitemOpen
  \href@noop {} {}\bibinfo {note} {Although the harmonic approximation is
  obviously better for larger $s$ values, we have checked that the dipolar tail
  calculated using Eq.~\eqref{eq:function-f} is in excellent agreement with
  that evaluated using the exact Wannier function down to $s=8$, see
  Fig.~\ref{fig:1}~(a). In particular, for $s=8$, the effective exponent
  $\beta_{\mathrm{eff}}$ evaluated from the harmonic approximation differs less
  than $3\%$ from the its exact value for the whole range of $\chi$
  considered.}\BibitemShut {Stop}%
\bibitem [{Gj()}]{Gj}%
  \BibitemOpen
  \href@noop {} {}\bibinfo {note} {We stress that $\beta_{\mathrm{eff}}$ only
  relates to $G_2$, and not to the overall decay $G_j$ of the dipolar tail,
  which is not generally described by a power law.}\BibitemShut {Stop}%
\end{thebibliography}
%

\appendix
\vspace{1cm}
\section{Derivation of dipolar tail}
\label{sec:App1}
The dipole-mediated inter-site interaction between particles separated by $j$ sites is characterized by the coupling constant:
\begin{equation}
V_j = \int d^3 r \int d^3r' V(\vec r-\vec r\,') |\varphi(\vec r)|^2 |\varphi(\vec r\,'-j\lambda \vec e_z)|^2,
\end{equation}
where we neglect the effect of exchange terms, since they are negligibly small for a sufficiently deep lattice. In the following calculation, we assume that the dipole 
is oriented along the lattice axis $z$ ($\alpha=0$ in the notation of the main text), but as discussed below the calculation can be easily generalized to any dipole orientation.

As discussed in the main text, we may approximate the on-site wavefunction by a Gaussian:
\begin{equation}
|\varphi(\vec r)|^2=\frac{e^{-z^2/l^2}}{\sqrt{\pi}l} \frac{e^{-\rho^2/l_\perp^2}}{\pi l_\perp^2},
\end{equation}
where $\rho = \sqrt{ x^2+y^2}$.
Then the Fourier transform of the density is of the form: $\tilde n_j(\vec k)=\tilde n(\vec k) e^{ik_z j \lambda}$, and 
\begin{equation}
\tilde n(\vec k)=e^{-k_z^2 l^2/4}e^{-k_\rho^2 l_\perp^2/4}.
\end{equation}
Using the convolution theorem we may then re-express $V_j$ in the form:
\begin{equation}
V_j \simeq \int \frac{d^3k}{(2\pi)^3} \tilde V(\vec k) \tilde n_0(\vec k) \tilde n_j(\vec k)
\end{equation}
with $\tilde V(\vec k)=\frac{4\pi d^2}{3}\left [ \frac{3k_z^2}{|\vec k|^2}-1\right ]$, the Fourier transform of the dipole-dipole interaction potential. 
Using the form of $\tilde n_j(\vec k)$, we may then re-write:
\begin{eqnarray}
V_j &\simeq& \frac{2d^2}{3\pi l l_\perp^2} \int_0^1 du \left [ \frac{3u^2}{\Lambda^2 + (1-\Lambda^2)u^2} - 1 \right ]  \nonumber \\
&\times& \int_0^\infty dq q^2 e^{-q^2/2}\cos\left ( q\frac{z_j}{l}u \right ).
\end{eqnarray}
with $\Lambda=l/l_\perp$. For the specific case of a harmonic confinement with frequency $\omega_\perp$ and an optical lattice along $z$ with depth $V_0$, we may define, as in the main text, $\chi=\frac{\hbar\omega_\perp}{E_{\mathrm{R}}}$ and $s=\frac{V_0}{E_{\mathrm{R}}}$, and re-express 
$\Lambda(\chi,s)=\frac{1}{\sqrt{2}}\sqrt{\chi}\frac{1}{s^{1/4}}$. 
We can then write:
\begin{equation}
V_j \simeq \frac{2d^2}{3\sqrt{2\pi}l l_\perp^2} F_j(\chi,s)
\end{equation}
with 
\begin{eqnarray}
F_j(\chi,s) &=& \int_0^1 du \left [ \frac{3u^2}{\Lambda^2 + (1-\Lambda^2)u^2} - 1 \right ] \nonumber \\
&\times& \left [ 1-\tilde z_j^2 u^2\right ]
e^{-\frac{1}{2}\tilde z_j^2 u^2}
\end{eqnarray}
with $\tilde z_j = j \pi s^{1/4}$. Changing the integration variable into $\tilde u = \tilde z_j u$, and re-organizing the integrand we 
can re-write the integral in the form:
\begin{widetext}
\begin{equation}
F_j(\chi,s) = \frac{1}{\tilde z_j}\int_0^{\tilde z_j} du \left [ 
\left ( \frac{2+\Lambda^2}{1-\Lambda^2} \right ) - \left (\frac{3}{1-\Lambda^2}\right ) \frac{1}{1+\frac{1-\Lambda^2}{\Lambda^2}\frac{\tilde u^2}{\tilde z_j^2}} \right ] (1-\tilde u^2)e^{-\tilde u^2/2}
\end{equation}
\end{widetext}
For a sufficiently large $s$, due to the rapidly decaying exponential we can safely move the integral boundary to infinity. Note that this is only possible if $\Lambda<1$~($l_\perp>l$). 
We can then perform the integral analytically, obtaining:
\begin{equation}
\frac{V_j}{E_{\mathrm{R}}} =\frac{3}{\pi^2}B^{3/2}\left ( \frac{a_{dd}}{\lambda} \right ) f(\sqrt{B}j)
\label{eq:Vj-a0}
\end{equation}
where $f(\xi)$  is the expression of Eq.~(4) of the main text. The procedure for other dipole orientations is identical, and we may obtain  the general expression of Eq. (3) of the main 
text.

As a side remark, we compare the result of Eq.~\eqref{eq:Vj-a0} and the known result for the interaction between two dipoles in a 1D system (in absence of lattice) when the dipole is oriented along the system axis~\cite{Sinha2007}.  We can re-write Eq.~\eqref{eq:Vj-a0} in the form:
\begin{equation}
\frac{V_j}{E_{\mathrm{R}}} =3\pi \left ( \frac{\chi}{2} \right )^{3/2} \left ( \frac{\tilde a_{dd}}{\lambda} \right )  f \left (  \tilde j \pi \sqrt{ \frac{\chi}{2} } \right ),
\label{eq:Vj-a0-2}
\end{equation}
with $\tilde a_{dd} = a_{dd}/\left (1-\frac{\chi}{2\sqrt{s}} \right )^{3/2}$ and $\tilde j= j/\sqrt{1-\frac{\chi}{2\sqrt{s}} }$. Comparing to the result of Ref.~\cite{Sinha2007}, we note that Eq.~\eqref{eq:Vj-a0-2} acquires the same form as the interaction between two dipoles with a regularized dipole length $\tilde a_{\mathrm{dd}}$ separated by an effective distance $\tilde j \lambda$. 

\section{Asymptotic expression for large $\Lambda$}
\label{sec:App2}
Let us consider at this point the case  $\Lambda=l/l_\perp \gg 1$, which is the typical case in strong 3D optical lattices. 
In that case, we may approximate for sufficiently deep lattices:
\begin{equation}
V_j = \frac{2d^2}{3\sqrt{2\pi}l l_\perp^2} \frac{3}{\Lambda^2}  \int_0^1 du \frac{u^2}{(1-u^2)}  
\left [ 1-\tilde z_j^2 u^2\right ]
e^{-\frac{1}{2}\tilde z_j^2 u^2}.
\end{equation}
Since only small $u$ contribute for large $s$, we may expand:
\begin{widetext}
\begin{eqnarray}
V_j &\simeq&  \frac{2d^2}{3\sqrt{2\pi}l l_\perp^2} \frac{3}{\Lambda^2} 
\int_0^1 du  \left ( u^2 +(1-\tilde z_j^2)u^4 + \cdots \right )
e^{-\frac{1}{2} \tilde z_j^2 u^2} \nonumber \\
&\simeq& \frac{2d^2}{3\sqrt{2\pi}l l_\perp^2} \frac{3}{\Lambda^2} \frac{\sqrt{\pi}}{2}
\left [ 
\frac{\sqrt{2}}{\tilde z_j^3} +(1-\tilde z_j^2) \frac{3\sqrt{2}}{\tilde z_j^5}+\cdots
\right ] \mathrm{erf}\left ( \frac{\tilde z }{\sqrt{2}}\right ) \nonumber \\
&=& \frac{-2d^2}{\lambda^3}\frac{1}{j^3}\left [1+\frac{6}{j^2\pi^2\sqrt{s}} +  \cdots \right ]
\end{eqnarray}
\end{widetext}

Then 
\begin{equation}
\frac{V}{V_2} \simeq 8 \left [1+\frac{9}{2\pi^2\sqrt{s}} + \cdots\right ] 
\end{equation}
Now, writing $\frac{V}{V_2}=2^{\beta_{\mathrm{eff}}}$, and anticipating that $\beta_{\mathrm{eff}}-3 \ll 1$, we may expand:
\begin{equation}
2^{\beta_{\mathrm{eff}}} \simeq 8 \left [1+\mathrm{ln}(2) (\beta_{\mathrm{eff}}-3) \right ]
\end{equation}
Comparing both expressions we get the final result:
\begin{equation}
\beta_\infty \simeq 3 + \frac{1}{\mathrm{ln} (2)}  \frac{9}{ 2\pi^2  \sqrt{s}} 
\end{equation}
written in the main text.

\section{Time-Dependent Variational Principle algorithm implementation}
\label{sec:TDVP}
The Time-Dependent Variational Principle~(TDVP) algorithm~\cite{Haegeman11, Koffel12, Haegeman16, Goto19} may allow for studying the time evolution, for a limited time, of systems formed by hundreds of sites. In the algorithm, the time evolved state is represented as a matrix product state (MPS) (for a review see, e.g.~\cite{Schollwoeck11}). Such states are represented by tensors, which, apart from physical indices (site number, Hilbert space dimension on a given site) are characterized, at each site, by an auxiliary index running over a space of dimension $\chi$. The higher $\chi$, the more entangled states can be typically faithfully represented by the MPS representation.  During the time evolution with a many-body Hamiltonian, the entanglement in the time-evolved state typically grows with a rate dependent on the properties of the system. The rapid growth of the entanglement of an initially low-entangled state prevents from evolving the state for too long, typically limiting evolutions beyond tens of tunneling times in the ergodic regime. 
Once the motion becomes non-ergodic, and in particular localized, the growth of entanglement in time is much slower, which allows for faithfully tracing the time evolution even up to times of the order of thousand tunnelling times (for recent implementations for short range Hamiltonians see \cite{Chanda19,Chanda20,Chanda20m,Chanda20c,Sierant22}, where all the details are discussed). We use the same implementation, extended to the dipolar-like long-range terms using the so called matrix-product-operator representation of the Hamiltonian, implemented within the Itensor library~\cite{itensor}.



\begin{figure} [t]
\begin{center}
\includegraphics[width=\columnwidth]{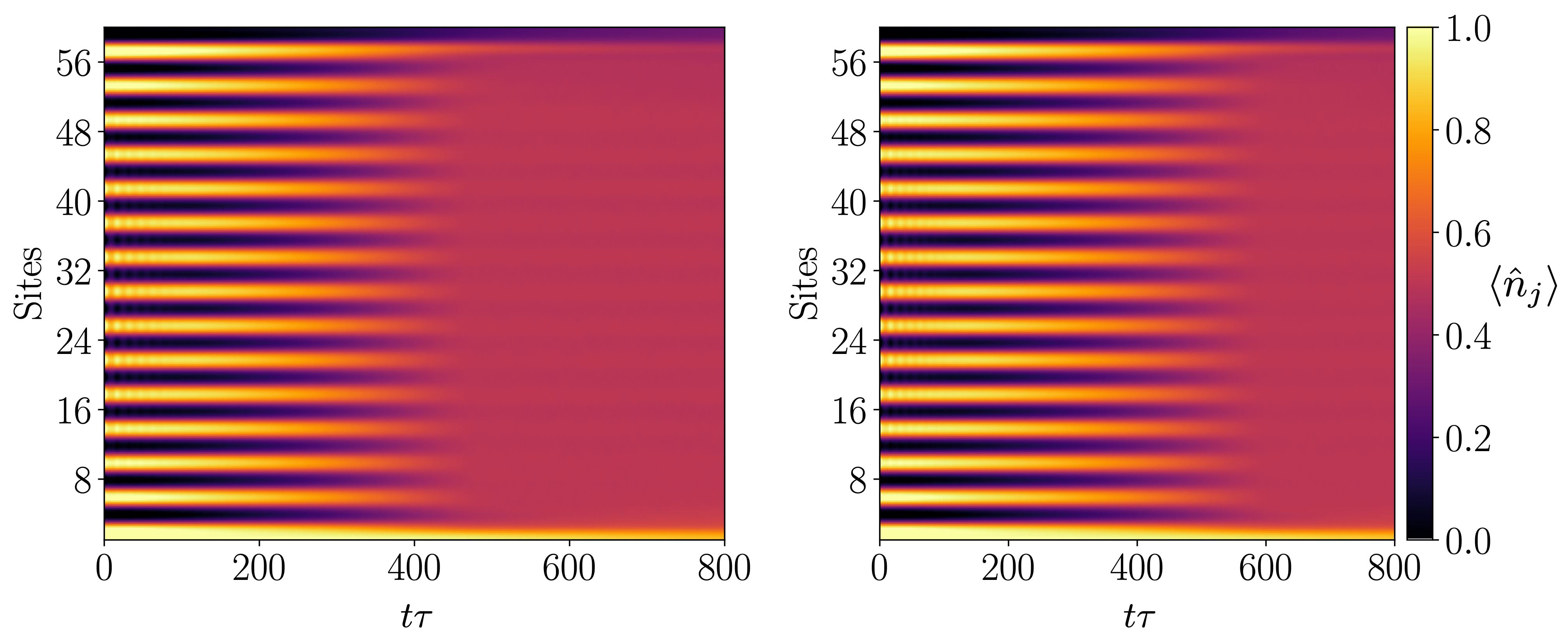}
\caption{Homogenization plot for $B=2.54$ $V/t=16$ obtained with TDVP for $\chi=256$ (left) and $\chi=384$ (right). Observe that more accurate results for larger $\chi$ value undergo homogenization in longer times. This is a typical behavior for not fully converged TDVP algorithm~\cite{Chanda19}. }
 \label{fig:sub}
\end{center}
\end{figure}


%

Due to the long range coupling as well as to the large value of $V/t$, the algorithm requires a large amount of CPU time for propagation. For the data presented in Fig.~3 of the main text, the propagation took more than a month on a single thread of a fast workstation for $\chi=384$~(less than a week for $\chi=256$). While the results 
were slightly different for $\chi=256$ and $384$~(see Fig.~\ref{fig:sub}), indicating lack of convergence in the delocalized regime, the qualitative time-dependence obtained was the same, with homogenization~(delocalization) occuring slightly slower for a larger (more accurate) $\chi$ value, confirming the claims expressed in the main text.



\begin{figure} [t]
\begin{center}
\includegraphics[width=\columnwidth]{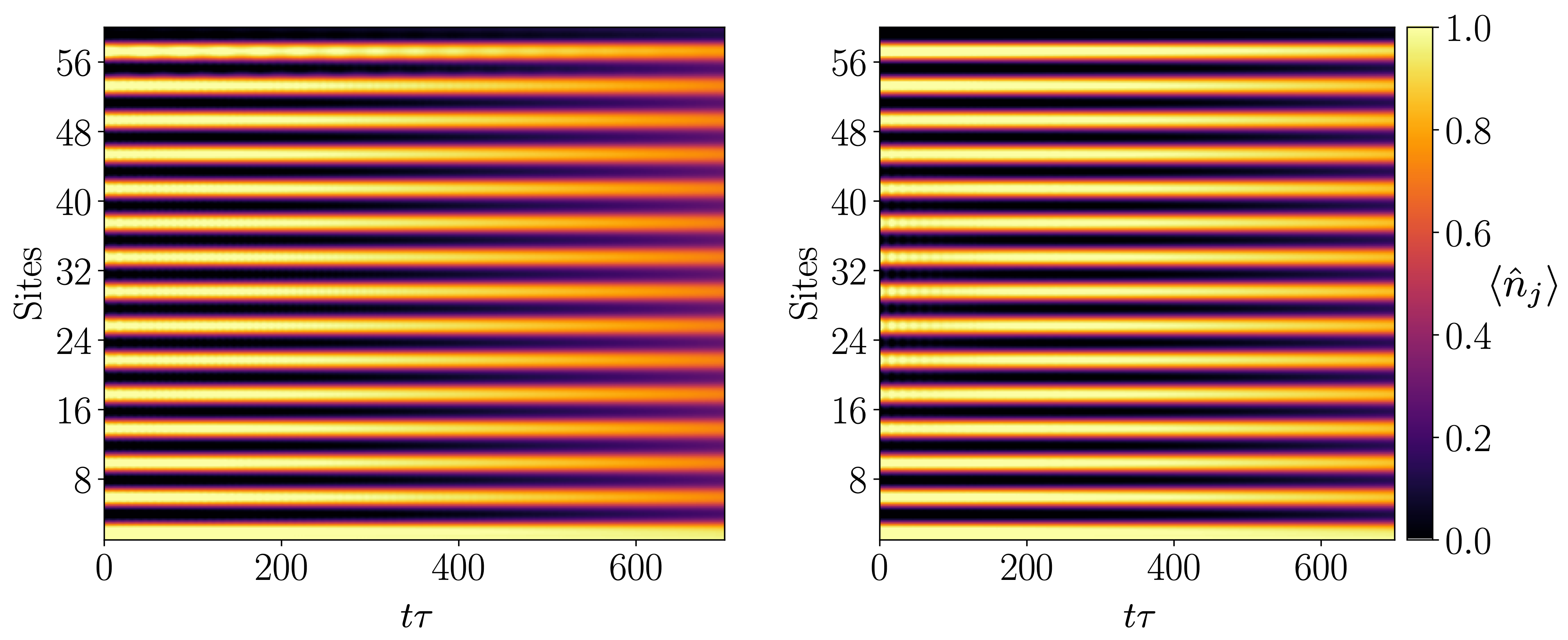}
\caption{
Homogenization plot for $V/t=20$ obtained with TDVP for $\chi=384$, for the $1/j^3$ model~(left) and for $G_j(B)$ with $B=2.54$~(right). 
Whereas the system remains fully localized in the latter model, 
the $1/j^3$ model shows the start of the density homogenization, which fully develop at later times.
}
 \label{fig:sub2}
\end{center}
\end{figure}


Figure~\ref{fig:sub2} shows the exemplary dynamics 
obtained for larger value of $V/t$ with converged results up to time $t=700/t$ for both the $1/j^3$ model and the case with 
the $G_j(B)$ tail. As in Fig.~\ref{fig:3}, the $1/j^3$ model 
leads eventually to deloclaization at later times, whereas the model with the correct dipolar tail shows a fully localized dynamics. \\

\section{Chebyshev propagation implementation}
\label{App:4}
We use the Chebyshev propagation scheme, as described in detail in~\cite{Fehske08} for $L=24$ at half filling. In this approach, the time evolution operator 
$U( \Delta t) = \exp(-i H \Delta t)$ over time period $\Delta t$ is expanded as:
\begin{equation}
 U( \Delta t) \approx \mathrm{e}^{-\mathrm{i}b \Delta t} \left( J_0(a \Delta t) + 2\sum_{k=1}^N (-i)^k J_k(a \Delta t) T_k \left( \mathcal{H} \right) \right),
 \label{eqcheby}
\end{equation}
where $a=(E_{\rm max} - E_{\rm min})/2$, $b=(E_{\rm max} + E_{\rm min})/2$ and $E_{\rm min}$/$E_{\rm max}$ is the lowest/highest eigenstate energy of the Hamiltonian $H$.  $\mathcal{H} = \frac{1}{a}(H-b)$ is the rescaled Hamiltonian with the spectrum in the $[-1,1]$ interval,
$J_k(t)$ is the Bessel function of the order $k$ and $T_k(x)$ is the Chebyshev polynomial of order $k$.
The order of the expansion $N$ and the time step $\Delta t$ is controlled by the preservation of the unitarity of the evolution allowing for numerically exact results, for further details see Ref.~\cite{Sierant22}.

\end{document}